\documentclass[twocolumn,showpacs,amsmath,amssymb,nofootinbib,superscriptaddress]{
revtex4}

\usepackage{graphicx}
\usepackage{epsfig}
\usepackage{bm}
\usepackage{amsfonts}
\usepackage[latin1]{inputenc}
\usepackage{amssymb}
\usepackage{color}
\usepackage{float}
\usepackage{amsmath}                  
\usepackage{dcolumn}
\usepackage{hyperref}
\usepackage{array}
\usepackage{lscape}

\usepackage{array}
\usepackage{booktabs}
\usepackage{multirow}

\providecommand{\U}[1]{\protect\rule{.1in}{.1in}}

\def\nn{\nonumber}

\def\be{\begin{equation}}
\def\ee{\end{equation}}
\def\bea{\begin{eqnarray}}
\def\eea{\end{eqnarray}}
\def\eqi{\begin{equation}}
\def\eqf{\end{equation}}
\def\eqia{\begin{eqnarray}}
\def\eqfa{\end{eqnarray}}

\def\a{\alpha}
\def\b{\beta}

\def\d{\delta}

\def\la{\lambda}

\def\k{\kappa}
\def\m{\mu}
\def\n{\nu}

\def\f{\phi}

\begin{document}

\title{Bounce and cyclic cosmology in new gravitational scalar-tensor theories}

\author{Emmanuel N. Saridakis}
\email{Emmanuel\_Saridakis@baylor.edu}
\affiliation{Chongqing University of Posts \& Telecommunications, Chongqing, 400065, 
China}
\affiliation{Department of Physics, National Technical University of Athens, Zografou
Campus GR 157 73, Athens, Greece}
\affiliation{ Eurasian  International Center for Theoretical
Physics, Eurasian National University, Astana 010008, Kazakhstan}

\author{Shreya Banerjee }\email{shreya.banerjee@tifr.res.in}
\affiliation{Tata Institute of Fundamental 
Research, Homi Bhabha Road, Mumbai 400005, India}

\author{R. Myrzakulov}\email{rmyrzakulov@gmail.com}
\affiliation{ Eurasian  International Center for Theoretical
Physics, Eurasian National University, Astana 010008, Kazakhstan}


\begin{abstract}
We study  the bounce and cyclicity realization in the framework of new gravitational 
scalar-tensor theories. In  these theories the Lagrangian contains the Ricci scalar and 
its first and second derivatives, in a specific combination that makes them free of 
ghosts, and transformed into the Einstein frame they are proved to be a subclass of 
bi-scalar extensions of general relativity. We present analytical expressions for the 
bounce requirements, and we examine the necessary  qualitative behavior of the 
involved functions that can give rise to a given scale factor. Having in mind  
these qualitative forms, we reverse the procedure and  we construct 
suitable simple Lagrangian functions that can give rise to a bounce or cyclic scale 
factor.  
\end{abstract}

\maketitle

\section{Introduction}

Although inflation is considered to be an important part of the universe history  
\cite{inflation}, the ``problem of the initial singularity'' is still present in the 
standard model of the universe. In particular, since such a singularity is unavoidable 
if inflation is driven by a scalar field in the framework of general 
relativity \cite{Borde:1993xh},  a lot of effort has been devoted in resolving it through 
quantum gravity considerations or effective field 
theory applications.

Non-singular bouncing cosmologies may offer a potential solution  to the cosmological 
singularity problem \cite{Mukhanov:1991zn}. Modified gravities are an ideal framework for 
their realization, since they allow for the necessary violation of the null energy 
condition \cite{Nojiri:2006ri,Capozziello:2011et}. In particular, one can obtain bouncing 
solutions in Pre-Big-Bang \cite{Veneziano:1991ek} 
and  Ekpyrotic \cite{Khoury:2001wf,Khoury:2001bz} models, $f(R)$ gravity 
\cite{Bamba:2013fha,Nojiri:2014zqa}, $f(T)$ 
gravity \cite{Cai:2011tc},
gravity actions with higher 
order corrections \cite{Tirtho1,Nojiri:2013ru},  braneworld scenarios 
\cite{Shtanov:2002mb,Saridakis:2007cf}, 
non-relativistic gravity \cite{Cai:2009in,Saridakis:2009bv}, Lagrange modified gravity 
\cite{Cai:2010zma}, massive gravity  
\cite{Cai:2012ag},  loop quantum cosmology 
\cite{Bojowald:2001xe,Odintsov:2014gea,Odintsov:2015uca} 
etc. Alternatively, bouncing cosmology can be realized   introducing matter 
fields that violate the null energy condition 
\cite{Cai:2007qw,Cai:2009zp,Nojiri:2015fia}, or constructing non-conventional 
mixing terms \cite{Saridakis:2009jq,Saridakis:2009uk}.
Furthermore, one may extend bouncing cosmology to the  paradigm of cyclic cosmology 
\cite{tolman}, in which the universe 
experiences a sequence of expansions and 
contractions \cite{Steinhardt:2001st,Steinhardt:2002ih} (see \cite{Novello:2008ra} for a 
review). This offers alternative insights for the origin of the observable universe 
\cite{Lidsey:2004ef,cyclic1,Nojiri:2011kd}, and can explain the
scale invariant power spectrum \cite{Novello:2008ra,Finelli:2001sr} and possible 
non-Gaussianities \cite{Cai:2009fn}. 

One recently constructed class of modified gravity is the so-called ``new gravitational 
scalar-tensor theories''  \cite{Naruko:2015zze, Saridakis:2016ahq}. In these theories one 
uses a Lagrangian with the Ricci scalar and its first and second derivatives, however in 
a specific combination that makes the theory free of ghosts. Transforming into the 
Einstein frame, one can show that these constructions propagate  $2+2$ degrees of 
freedom, and thus they fall outside Horndeski \cite{Horndeski:1974wa}, 
Galileon \cite{Nicolis:2008in,Deffayet:2009wt} and beyond Horndeski theories 
\cite{Gleyzes:2014dya}. Nevertheless, although these theories can be seen as a subclass of 
bi-scalar extensions of general relativity, they can still be expressed in pure 
geometrical terms, and that is why the authors called them ``new gravitational 
scalar-tensor theories''.  Due to the presence of extra degrees of freedom, these 
theories can lead to very interesting cosmological behavior 
\cite{Saridakis:2016ahq}.

In the present work, we are interested in studying the realization of bounce and 
cyclicity in the framework of new gravitational scalar-tensor theories.  The plan of the 
work is as follows: In Section \ref{themodel} we review the new gravitational 
scalar-tensor theories, and we apply them in a cosmological framework.
 In Section \ref{bouncycl} we construct specific subclasses of Lagrangian functions that 
give rise to bouncing and cyclic scale factors. Finally,  we summarize our results in 
section \ref{Conclusions}.

\section{New gravitational scalar-tensor theories}
\label{themodel}

In this section we briefly review the gravitational theories that include higher 
derivative curvature terms, which were named new gravitational scalar-tensor 
theories in \cite{Naruko:2015zze}. Transformed into the Einstein frame these theories 
present two extra scalar degrees of freedom comparing to general relativity, and thus 
they fall in the class of bi-scalar modifications.

The action of the new gravitational scalar-tensor theories is  
\begin{equation}
\label{bfR}
S=\int d^{4}\sqrt{-g}\, f\left(R,(\nabla R)^{2},\square R \right),
\end{equation}
with $(\nabla R)^{2}=g^{\m\n}\nabla_{\m}R\nabla_{\n}R$, and where for simplicity, here 
and in the following, we set the Planck mass $M_{pl}$ to one.  
Despite the presence of higher derivatives, these actions are ghost free and can be 
transformed into bi-scalar theories in the Einstein frame through double Lagrange 
multipliers. Although one can consider models, namely $f$-forms, nonlinear in  $\square 
R=\b$, in the present work we focus on theories with \cite{Saridakis:2016ahq}
\begin{equation}
f(R,(\nabla R)^{2},\square R)=\mathcal{K}((R,(\nabla R)^{2})+\mathcal{G}(R,(\nabla 
R)^{2})\square R.
\end{equation}
In this case, action (\ref{bfR}) can be transformed  to  
\begin{eqnarray}
\label{action}
&&\!\!\!\!\!\!\!\!\!\!\!\!\!\!\!
S=\int d^{4}x 
\sqrt{-\hat{g}}\left[\frac{1}{2}\hat{R}-\frac{1}{2}\hat{g}^{\m\n}\nabla_{\m}\chi\nabla_{\n
}\chi\right.\nonumber\\
&& \ \ \ \ \ \ \ \ \ \ \ \
- 
\frac{1}{\sqrt{6}}e^{-\sqrt{\frac{2}{3}}\chi}\hat{g}^{\m\n}\mathcal{G}\nabla_{\m}
\chi\nabla_{\n}\phi+\frac{1}{4}e^{
-2\sqrt{
\frac{2}{3}}
\chi}\mathcal{K}
\nonumber\\
&& \ \ \ \ \ \ \ \ \ \ \ \
\left.
+\frac{1}{2}e^{-\sqrt{\frac{2}{3}}\chi}\mathcal{G}\hat{\square}\phi-\frac{1}{4}e^{-\sqrt{
\frac{2}{3}}\chi}\phi\right],
\end{eqnarray}
where  
\be
\mathcal{K}=\mathcal{K}(\phi,B),\quad\mathcal{G}=\mathcal{G}(\phi,B),
\ee
 with
\be
B=2e^{\sqrt{
\frac{2}{3}}\chi}g^{\m\n}\nabla_{\m}\phi\nabla_{\n}\phi.
\ee
In the above expressions we have introduced the $\chi$-field through the conformal 
transformation $g_{\mu\nu}=\frac{1}{2}e^{-\sqrt{\frac{2}{3}}\chi}\hat{g}_{\mu\nu}$ 
(the $hat$ denotes the conformally related frame), while the $\f$-field enters through   
$\varphi\equiv f_{\b}$.
We mention here that the above action  contains two scalar fields, i.e. $\chi$ and 
$\f$, however in the specific combination that makes it equivalent to the original 
higher-derivative gravitational action. Thus, these theories consist pure gravitational 
formulations of standard multi-scalar-tensor theories. 

 We will work in the    Einstein-frame version of the above 
theories, namely with action (\ref{action}), and  for simplicity we neglect the hats.
Variation of (\ref{action}) with respect to the metric leads to  the  field 
equations 
\begin{eqnarray}
\label{metricfieldeq}
\mathcal{E}_{\m\n}&\!\!=\!\!&\frac{1}{2}G_{\m\n}+\frac{1}{4}g_{\m\n}g^{\a\b}\nabla_{\a}
\chi\nabla_
{\b}\chi-\frac{1}{2}\nabla_{\m}\chi\nabla_{\n}\chi\nonumber\\
&&
+\frac{1}{4}g_{\m\n}\sqrt{\frac{2}{3}}e^
{-\sqrt{ \frac{2}{3}}
\chi}g^{\a\b}\mathcal{G}\nabla_{\a}\chi\nabla_{\b}\phi\nonumber\\
&&
-\frac{1}{2}\sqrt{\frac{2}{3}}e^{
-\sqrt{\frac{
2}{3}}\chi}\mathcal{G}\nabla_{(\m}\chi\nabla_{\n)}\phi \nn\\
&&-\sqrt{\frac{2}{3}}g^{\a\b}\nabla_{\a}\chi\nabla_{\b}\phi\,\mathcal{G}_{B}\nabla_{\m}
\phi\nabla_{\n}\phi
\nonumber\\
&&
-\frac{1}{4}g_{\m\n}e^{-\sqrt{\frac{2}{3}}\chi}\mathcal{G}
\square\phi+\mathcal{G}_{B 
}(\square\phi)\nabla_{\m}\phi\nabla_{\n}\phi
\nonumber\\
&&
+\frac{1}{2}e^{-\sqrt{\frac{2}{3}}\chi}
\mathcal{G} \nabla_{\m}\nabla_{\n}\phi 
-\frac{1}{2}\nabla_{\k}\left(e^{-\sqrt{\frac{2}{3}}\chi}\mathcal{G}\d^{\la}_{(\m}\d^{\k}
_{\n)} 
\nabla_{\la}\phi\right)
\nonumber\\
&&
+\frac{1}{4}\nabla_{\k}\left(e^{-\sqrt{\frac{2}{3}}\chi}\mathcal{G}
g_{\m\n}\nabla^{\k}\phi\right)-\frac{1}{8}g_{\m\n}e^{-2\sqrt{\frac{2}{3}}\chi}\mathcal{K}
\nonumber\\
&&
+\frac{1}{ 2}e^{-\sqrt{\frac{2}{3}}\chi}\mathcal{K}_{B}\nabla_{\m}\phi\nabla_{\n}\phi 
+\frac{1}{8}g_{\m\n}e^{-\sqrt{\frac{2}{3}}\chi}\phi=0,
\end{eqnarray}
where the parentheses in spacetime indices denote symmetrization, and the subscripts in 
$\mathcal{G}$ and $\mathcal{K}$  mark partial derivatives   (e.g.
$\mathcal{G}_{B}=\frac{\partial \mathcal{G}(\phi,B)}{\partial B}$
etc). Furthermore, variation of (\ref{action}) with respect to  $\chi$ 
and $\phi$ gives rise to their equations of motion, namely
\begin{eqnarray}
\mathcal{E}_{\chi}&\!\!=\!\!&\square\chi+\frac{1}{3}e^{-\sqrt{\frac{2}{3}}\chi}g^{\m\n}
\mathcal{ G}
\nabla_{\m}\chi\nabla_{\n}\phi\nonumber\\
&&
-\frac{2}{3}g^{\m\n}\nabla_{\m}\chi\nabla_{\n}\phi\,\mathcal
{G}_{B}g^ 
{\a\b}\nabla_{\a}\phi\nabla_{\b}\phi\nonumber\\
&&
+\frac{1}{2}\sqrt{\frac{2}{3}}\nabla_{\m}\left( 
e^{-\sqrt{\frac{2}{3}}\chi}g^{\m\n}\mathcal{G}\nabla_{\n}\phi\right)\nn\\
&&-\frac{1}{2}\sqrt{\frac{2}{3}}e^{-\sqrt{\frac{2}{3}}\chi}\mathcal{G}\square\phi+\sqrt{
\frac{2}{3}}
\mathcal{G}_{B}\nabla_{\m}\phi\nabla_{\n}\phi\,g^{\m\n}\square\phi
\nonumber\\
&&
-\frac{1}{2}\sqrt{\frac{
2}{3}}e^{-
2\sqrt{\frac{2}{3}}\chi}\mathcal{K}+\frac{1}{2}e^{-\sqrt{\frac{2}{3}}\chi}\mathcal{K}_{B}
\sqrt{\frac{2}{3}}g^{\m\n}\nabla_{\m}\phi\nabla_{\n}\phi
\nn\\
&&+\frac{1}{4}\sqrt{\frac{2}{3}}e^{-\sqrt{\frac{2}{3}}\chi}\phi=0,
\label{chifieldeq}
\end{eqnarray}
and 
\begin{eqnarray}
\mathcal{E}_{\phi}&\!\!=\!\!&-\frac{1}{2}\sqrt{\frac{2}{3}}e^{-\sqrt{\frac{2}{3}}\chi}g^{
\m\n }
\mathcal{G}_{\phi}\nabla_{\m}\chi\nabla_{\n}\phi
\nonumber\\
&&
+2\sqrt{\frac{2}{3}}\nabla_{\b}\left(g^{
\m\n}\mathcal{G} 
_{B}g^{\a\b}\nabla_{\a}\phi\nabla_{\m}\chi\nabla_{\n}\phi\right)
\nonumber\\
&&
+\frac{1}{2}\sqrt{\frac{2}
{3}} 
\nabla_{\n}\left(e^{-\sqrt{\frac{2}{3}}\chi}g^{\m\n}\mathcal{G}\nabla_{\m}\chi\right)\nn\\
&&+\frac{1}{2}e^{-\sqrt{\frac{2}{3}}\chi}\mathcal{G}_{\phi}\square\phi-2\mathcal{G}_{B}
(\square\phi)
^{2}-2\nabla_{\n}\mathcal{G}_{B}\square\phi\nabla^{\n}\phi
\nonumber\\
&&
-\frac{1}{2}\sqrt{\frac{2}{3}}
\nabla^{\m}\left(e^{-\sqrt{\frac{2}{3}}\chi}\nabla_{\m}\chi\,\mathcal{G} 
\right)+\frac{1}{2}\nabla^{\m}\left(e^{
-\sqrt{\frac{2}{3}}\chi}\,\mathcal{G}_{\phi}\nabla_{\m}\phi\right)\nn\\
&&-\frac{1}{2}\sqrt{\frac{2}{3}}e^{-\sqrt{\frac{2}{3}}\chi}\nabla^{\m}\chi\mathcal{G}_{B}
\nabla_{\m}
B+\frac{1}{2}e^{-\sqrt{\frac{2}{3}}\chi}\nabla^{\m}\mathcal{G}_{B}\nabla_{\m}B
\nonumber\\
&&
+\sqrt{\frac{2}{3}}
e^{-\sqrt{
\frac{2}{3}}\chi}\mathcal{G}_{B}\nabla^{\m}\left(e^{\sqrt{\frac{2}{3}}
\chi}\nabla_{\m}\chi\nabla^{\n}\phi\nabla_{\n}\phi\right)\nn\\
&&
+2 e^{-\sqrt{\frac{2}{3}}\chi}\mathcal{G}_{B}  
\nabla^{\m}\left(e^{\sqrt{\frac{2}{3}}\chi}\nabla^{
\n}\phi\right)\nabla_{\m}\nabla_{\n}\phi
\nonumber\\
&&
+2\mathcal{G}_{B}R^{\m\n}\nabla_{\m}\phi\nabla_{\n
}\phi+\frac{1}{4}e^{-2\sqrt{\frac{2}{3}}\chi}\mathcal{K}_{\phi}\nn\\
&&-\nabla_{\n}\left(e^{-\sqrt{\frac{2}{3}}\chi}\mathcal{K}_{B}g^{\m\n}\nabla_{\m}
\phi\right)-\frac{
1}{4}e^{-\sqrt{\frac{2}{3}}\chi}=0.
\label{phifieldeq}
\end{eqnarray}
As mentioned above, we do verify that all field equations do not contain 
problematic higher-derivative terms, and thus  theory (\ref{bfR}) is indeed healthy as it is
constructed to be. Finally, note that in the scenario at hand  general relativity is
reproduced   when $\mathcal{K}=\phi/2$ and $\mathcal{G}=0$, and in this case the 
triviality of the conformal transformation gives
$\chi=-\sqrt{\frac{3}{2}}\ln2$.

We proceed by applying the above theories into a cosmological framework. We add the 
matter sector straightaway in the  Einstein frame and we consider the total action 
$S_{tot}=S+S_{m}$ \cite{Saridakis:2016ahq}. Therefore, the metric  
field equations (\ref{metricfieldeq})   become
\be
\mathcal{E}_{\m\n}=\frac{1}{2}T_{\m\n},
\ee
with $T_{\m\n}=\frac{-2}{\sqrt{-g}}\frac{\d S_{m}}{\d g^{\m\n}}$   the energy-momentum 
tensor of the matter sector considered to correspond to a perfect fluid. Moreover, we 
consider a flat 
Friedmann-Robertson-Walker (FRW) geometry with metric
 \be
ds^{2}=-dt^{2}+a(t)^{2} \delta_{ij}dx^{i}dx^{j},
\ee
with $a(t)$ is the scale factor, and hence the two scalars are time-dependent only. With 
these considerations, equations (\ref{metricfieldeq}) lead to the two
Friedmann equations:
\begin{eqnarray}
&&\!\!\!\!\!\!\!\!\!\!\!\!\!\!\!\!\!
3H^{2}-\rho_m-\frac{1}{2}\dot{\chi}^{2}+\frac{1}{4}e^{-2\sqrt{\frac{2}{
3}}\chi}\mathcal{K}
\nonumber\\
&& \!\!\!\!
+\frac{2}{3}\dot{\phi}^{2}\left[\dot{\phi}\left(\sqrt{6}\dot{\chi}
-9H\right)-3\ddot{\phi} 
\right]\mathcal{G}_{B}
\nn\\
&&  \!\!\!\!
-\frac{1}{2}e^{-\sqrt{\frac{2}{3}}\chi}\left[\dot{B}\dot{\f}\mathcal{
G}_{B}+ \frac{\f}{2} +\dot{\f}^{2}\left(\mathcal{G}_{\f}-2\mathcal{K}_{B} \right) 
 \right]\! =0,  
\label{FR1}
\end{eqnarray}
\begin{eqnarray}
&&\!\!\!\!\!\!\!\!\!\!\!\!\!\!\!\!\!\!\!\!\!\!
3H^{2}+2\dot{H}+p_m+\frac{1}{2}\dot{\chi}^{2}+\frac{1}{4}e^{-2\sqrt{
\frac{2}{3}}
\chi}\mathcal{K}
\nonumber\\
&&\!\!\!\!\!\!\!\!\!
+\frac{1}{2} e^{-\sqrt{\frac{2}{3}}\chi}\left(-\frac{\f}{2}+\dot{B}\dot{\f
}\mathcal{
G}_{B}+\dot{\f}^{2}\mathcal{G}_{\f} \right)=0,
  \label{FR2}
\end{eqnarray}
with $B(t)=2 e^{\sqrt{\frac{2}{3}}\chi} g^{\m\n}\nabla_{\m}\f\nabla_{\n}\f=-2 
e^{\sqrt{\frac{2}{3}}\chi} \dot{\f}^{2}$,  $H=\dot{a}/a$   the Hubble 
parameter, and where dots denote differentiation with respect to $t$. Additionally, we 
have introduced the energy density  $\rho_m$ and pressure $p_m$  of the matter fluid.
Similarly,  the two scalar field equations (\ref{chifieldeq}) and (\ref{phifieldeq}) 
lead to the scalar evolution equations:
\begin{eqnarray}
&&\!\!\!\!\!\!\!\!\!\!\!\!\!\!
\mathcal{E}_{\chi}=\ddot{\chi}+3H\dot{\chi}-\frac{1}{3}\dot{\f}^{2}\left[\dot{\f}
\left(3\sqrt{6}H-
2\dot{\chi} \right)+\sqrt{6}\ddot{\f} \right]\mathcal{G}_{B}\nn\\
&&+
\frac{1}{
2\sqrt{6}}
e^{-\sqrt{\frac{2}{3}}\chi}\left[2\dot{B}\dot{\f}\mathcal{G}_{B}-\f+2\dot{\f}^{2}
\left(\mathcal{K}_{
B}+\mathcal{G}_{\f} \right) \right]\nonumber\\
&&+\frac{1}{\sqrt{6}}e^{-2\sqrt{\frac{2}{3}}\chi}\mathcal{K}
=0,
\label{chiequation}
\end{eqnarray}
and
\begin{eqnarray}
&&\!\!\!\!\!\!\!\!\! \!
\mathcal{E}_{\phi}=
\frac{1}{3}e^{-\sqrt{\frac{2}{3}}\chi}
\left[\dot{\f}\left(-9H+\sqrt{6}\dot{\chi}
\right)-3\ddot{\f}\right]\mathcal{K}_{B}
\nn
\\
&&
+\frac{1}{6}\dot{B}\left\{3e^{-\sqrt{\frac{2}{3}}
\chi}\dot{B} +4\dot{\f}
\left[\dot{\f}\left(9H-\sqrt{6}\dot{\chi}\right)+3\ddot{\f}\right]\right\}\mathcal{G}_{BB}
\nn\\
&&
+\frac{1}{3}e^{-\sqrt{
\frac{2}{3}
}\chi}\left[\dot{\f}\left(9H-\sqrt{6}\dot{\chi}\right)+3\ddot{\f}\right]\mathcal{G}_{\f}
\nn
\\
&&
+\left\{e^{-\sqrt{\frac{2}{3}}\chi}\dot{B}\dot{\f}+\frac{2}{3}\dot{\f}^{2}
\left[\dot{\f}\left(9H-\sqrt{6}\dot{\chi}\right)+3\ddot{\f}\right]\right\}\mathcal{G}_{B 
\f}
\nn
\\
&&
-e^{-\sqrt{\frac{2}{3}}\chi}\dot{\f}^{2}\mathcal{K}_{B 
\f}+\frac{1}{2}e^{-\sqrt{\frac{2}{3}}\chi}\dot{\f}^{2}\mathcal{G}_{\f\f}
 -e^{-\sqrt{\frac{2}{3}}\chi}\dot{B}\dot{\f}\mathcal{K}_{BB}
 \nn
\\
&&
+\left[
 \frac{4}{3}\dot{\f}\left(9H-2\sqrt{6}\dot
{\chi} 
\right)\ddot{\f}
-\frac{1}{\sqrt{6}}e^{
-\sqrt{\frac{2}{3}}\chi}\dot{B}\dot{\chi}
\right.\nn\\
&&\left. \ \ \ \,
+\dot{\f}^2\left(18H^{2}+6\dot{H}-3\sqrt{6}H\dot{\chi}-\frac{2}{3}\dot{\chi}^{2}
-\sqrt{6} \ddot{\chi}\right)\right]
\mathcal{G}_{B}
\nn
\\
&&
-\frac{1}{4}
e^{-2\sqrt{\frac{2}{3}}\chi}\mathcal{K}_{\f}+ \frac{1}{4}e^{-\sqrt{\frac{2}{3}}\chi}
=0,
\label{phiequation}
\end{eqnarray}
with   $\mathcal{G}_{B \f}=\mathcal{G}_{\f 
B}\equiv\frac{\partial^{2}\mathcal{G}}{\partial B \partial \f}$, etc.  
We mention that amongst the above four equations, namely (\ref{FR1}),(\ref{FR2}),
(\ref{chiequation}),(\ref{phiequation}), only three are independent, due to the fact that 
the total action is diffeomorphism  invariant, i.e. we have a conservation equation
\cite{Saridakis:2016ahq}:
\be
\nabla_\m \mathcal{E}^{\m\n}+\frac{1}{2}\mathcal{E}_{\chi} 
\nabla^{\n}\chi+\frac{1}{2}\mathcal{E}_{\f} 
\nabla^{\n}\phi=\frac{1}{2}\nabla_{\m}T^{\m\n}=0.
\ee 
Thus, the matter energy density and pressure satisfy the standard
evolution equation 
\begin{equation}
\label{rhoevol}
\dot{\rho}_m+3H(\rho_m+p_m)=0.
\end{equation}

The investigation of the above cosmological scenario at late times was performed in 
\cite{Saridakis:2016ahq}, where it was indeed shown that one can obtain interesting 
phenomenology. In the present work we are interested in studying the early-time phases, 
and in particular to examine whether the bounce realization is possible. This is 
performed in the following sections.

\section{Bouncing and cyclic solutions}
\label{bouncycl}

In this section we proceed to the investigation of  bounce and cyclicity realization 
in cosmology driven by new gravitational scalar-tensor theories. As it is known, in order 
to obtain a bounce we need a contracting phase ($H<0$) succeeded by an expanding one  
($H>0$), and thus at the bounce point we have $H=0$ and $\dot H > 0$. On the other hand, for the 
turnaround realization we need the succession of an expanding and a contracting phase, 
and 
thus at the turnaround point  $H=0$ and $\dot H<0$. Although these conditions cannot be 
fulfilled in the framework of general relativity \cite{Nojiri:2013ru},  observing the form 
of the two Friedmann equations 
(\ref{FR1}),(\ref{FR2}), as well as of the  scalar-field equations
(\ref{chiequation}),(\ref{phiequation}) we 
deduce that for suitable choices of the free functions $\mathcal{K}$ and 
$\mathcal{G}$  one can obtain the necessary violation of the null energy condition 
and thus satisfy the aforementioned bouncing and cyclic conditions.

\subsection{Reconstruction of a bounce}
\label{reconstructionb}

Let us now proceed to the investigation of the bounce realization. Suppose 
that we impose a given form of a bouncing scale factor, in which case   
  $H(t)$ and $\dot{H}(t)$ are also known. Substitution of this bouncing scale factor 
into the three independent equations (\ref{FR1}),(\ref{chiequation}) and 
(\ref{phiequation}), and recalling that in FRW geometry $B(t)= -2 
e^{\sqrt{\frac{2}{3}}\chi(t)} \dot{\f}(t)^{2}$, we obtain a system of three 
differential equations for the four functions $\phi(t)$ and $\chi(t)$ and for 
$\mathcal{K}$ and 
$\mathcal{G}$ (and their derivatives) considered as functions of $t$. Thus, we have the 
freedom to further consider the form of one of $\mathcal{K}$ and 
$\mathcal{G}$. In the following paragraphs we examine two such cases separately, having 
in mind that in order to be able to obtain a bounce we need to go beyond the simple 
$\mathcal{K}$ and $\mathcal{G}$ forms investigated in \cite{Saridakis:2016ahq}, which 
were adequate to describe the late-time universe.

\subsubsection{ Model I: $\mathcal{K}=\phi/2$ and $\mathcal{G}=\mathcal{G}(B)$}

Since general relativity is re-obtained for  $\mathcal{K}=\phi/2$ and $\mathcal{G}=0$, 
one   class of viable models of new gravitational scalar-tensor theories is the one with
\begin{equation}
\mathcal{K}=\frac{\phi}{2},
\end{equation}
and with 
\begin{equation}
\mathcal{G}=\mathcal{G}(B),
\end{equation}
i.e. $\mathcal{G}$ is  independent of $\phi$. Concerning the explicit bouncing scale 
factor, without loss of 
generality we consider the matter bounce form \cite{Cai:2011tc}
\begin{equation}
a(t)=a_b(1+qt^2)^{1/3},
\label{matterbounce}
\end{equation}
with $a_b$ denoting the value of scale factor at the bounce point $t=0$, and 
$q$ a positive parameter which determines how fast the
bounce is realized. From this scale factor we immediately find
\begin{eqnarray}
\label{Htt}
&&H(t)=\frac{2qt}{3(1+qt^2)}\\
&&\dot{H}(t)=\frac{2q}{3}\left[\frac{1-qt^2}{(1+qt^2)^2}\right].
\label{Hdottt}
\end{eqnarray}

Inserting these  considerations into equations
(\ref{FR1}),(\ref{chiequation}) and (\ref{phiequation}), 
and replacing 
$\mathcal{G}_B(t)=\mathcal{G}'(t)/B'(t)$, and 
$\mathcal{G}_{BB}(t)=(\mathcal{G}''(t)B'(t)-B''(t)\mathcal{G}')/(B'(t))^3$,
we obtain a system of differential 
equations for $\phi(t)$, $\chi(t)$ and $G(t)$. Unfortunately, this system cannot be 
solved analytically, however it can easily be elaborated numerically, leading to the 
extraction of the $\phi(t)$, $\chi(t)$ and $G(t)$. Since $B(t)$ can then be found, we can 
acquire the function $\mathcal{G}(B)$ in a parametric form. Hence, it is this 
reconstructed $\mathcal{G}(B)$ that generates the input scale factor (\ref{matterbounce}).

\begin{figure}[ht]
\centering
\includegraphics[width=8.4cm,height=5.4cm]{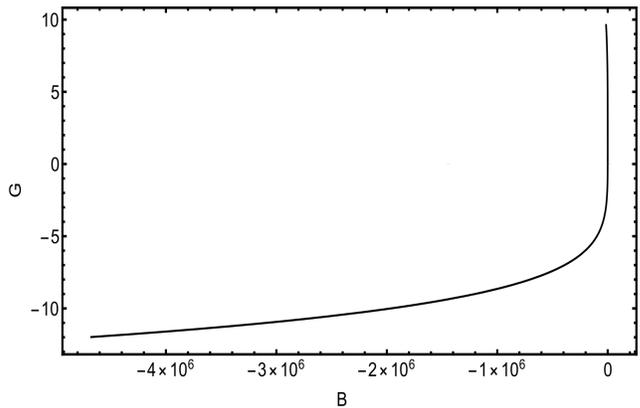}
\caption{   {\em The reconstructed $\mathcal{G}(B)$ that generates the bouncing scale 
factor (\ref{matterbounce}), in the case where 
$\mathcal{K}=\phi/2$. The bouncing parameters have been chosen as  
$ a_b=0.2$, $q=10^{-5}$. All quantities are measured in $M_{pl}$ units. 
}  }
\label{GB}
\end{figure}

In Fig. \ref{GB} we present the $\mathcal{G}(B)$ that is reconstructed from the 
given bouncing scale-factor form (\ref{matterbounce}), according to the above 
procedure, where we have neglected the matter sector in order to investigate the pure 
effect of the novel terms of the present theory. 
As we observe from the above procedure, and in particular from  Fig. \ref{GB},
in order to obtain a bouncing scale factor in the 
case where  $\mathcal{K}=\phi/2$, we need a $\mathcal{G}(B)$ function that resembles 
an exponential function of $B$. 

\begin{figure}[ht]
\centering
\includegraphics[width=8cm,height=5cm]{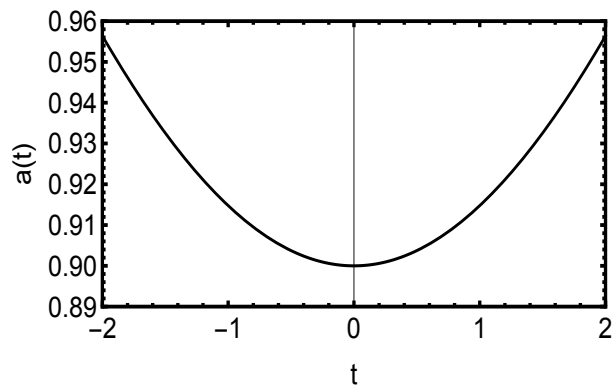}
\caption{  {\em The evolution of the scale factor $a(t)$ that is generated by the 
the exponential form $\mathcal{G}(B)= G_0 e^{G_1 B}$, in the case where 
$\mathcal{K}=\phi/2$, with $G_0=-7$ and $G_1=0.0001$ in $M_{pl}$ 
units. 
}  }
\label{scale1}
\end{figure}

The explicit example of the above reconstruction procedure offered us qualitative 
information for the $\mathcal{G}(B)$ form that leads to a bouncing scale factor.
Thus, we can now reverse the reconstruction procedure and impose the form of 
$\mathcal{G}(B)$ a priori, and then extract the induced scale factor, which is the 
physical procedure. Having the  qualitative requirements for  $\mathcal{G}(B)$ in mind, 
we choose its form to be
\begin{equation}
\mathcal{G}(B)= G_0 e^{G_1 B},
\label{Gform1}
\end{equation}
where $G_0$ and $G_1$ are parameters. Substituting it into equations (\ref{FR1}), 
(\ref{FR2}) and (\ref{chiequation}), considering once again $\mathcal{K}=\phi/2$, we 
acquire three second order differential equations for  
$a(t)$, $\phi(t)$ and $\chi(t)$. Elaborating the system numerically we extract the scale 
factor, and we depict it in  Fig. \ref{scale1}.
Hence, we do verify that new gravitational scalar-tensor theories with   
$\mathcal{K}=\phi/2$ and $\mathcal{G}(B)=G_0 e^{G_1 B}$ lead to the realization of a 
cosmological bounce.

The above procedure can easily be repeated in the presence of the matter sector which 
gives rise to matter bounce. We find that the same   forms of the functions 
$\mathcal{K}$ and $\mathcal{G}$  in the presence of pressureless matter, i.e. with $p_m=0$ 
and  $\rho_m\propto a^3$, give rise to a matter bounce, namely to a scale-factor similar 
to that of Fig. \ref{scale1}.

\subsubsection{ Model II: $\mathcal{K}=\phi/2+f(B)$ and $\mathcal{G}=\xi B$}

In this paragraph we investigate the bounce realization in a different subclass of new 
gravitational scalar-tensor theories. In particular, we choose 
\begin{eqnarray}
&&\mathcal{K}=\frac{\phi}{2}+f(B)\nonumber\\
&&  \mathcal{G}=\xi B,
\end{eqnarray}
where $f(B)$ is an unknown function of $B$ and $\xi$ is a 
parameter. Similarly to the previous paragraph, firstly we consider the bounce scale 
factor (\ref{matterbounce}), in order to numerically reconstruct  $f(B)$ and acquire a 
qualitative picture of its form. Indeed, substituting these  into equations
(\ref{FR1}),(\ref{chiequation}) and (\ref{phiequation}), and replacing  
$\mathcal{K}_B=f'(t)/B'(t)$ and 
$\mathcal{K}_{BB}(t)=(f''(t)B'(t)-B''(t)f')/(B'(t))^3$,
we obtain a system of differential equations for $\phi(t)$, $\chi(t)$ and $f(t)$.
Since this system cannot be solved analytically, we elaborate it numerically and we  
extract $\phi(t)$, $\chi(t)$ and $f(t)$. Since $B(t)$ can then be found as $B(t)= -2 
e^{\sqrt{\frac{2}{3}}\chi(t)} \dot{\f}(t)^{2}$, we can finally
acquire the function $f(B)$ in a parametric form. Hence,   this 
reconstructed $f(B)$   generates the input scale factor (\ref{matterbounce}).

\begin{figure}[ht]
\centering
\includegraphics[width=8cm,height=5cm]{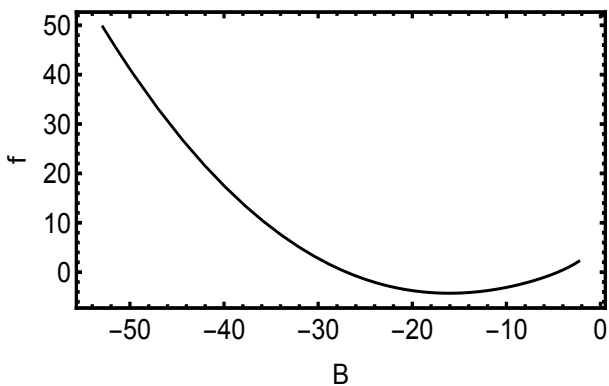}
\caption{\small  {\em The reconstructed $f(B)$  that generates the bouncing 
scale factor (\ref{matterbounce}), in the case where 
 $\mathcal{K}=\phi/2+f(B)$ and $\mathcal{G}=\xi B$. The bouncing parameters have been 
chosen as  
$ a_b=0.2$, $q=10^{-5}$, while $\xi=0.1$. All quantities are measured in $M_{pl}$ units.
}  }
\label{f(B)}
\end{figure}

In Fig. \ref{f(B)} we present the $f(B)$ that is reconstructed according to the above 
procedure, in the absence of matter sector. 
As we observe, in order to obtain a bouncing scale factor in the case where   
$\mathcal{K}=\phi/2+f(B)$ and $\mathcal{G}=\xi B$, we need a
$f(B)$ form that resembles a parabolic function of $B$.
 
\begin{figure}[ht]
\centering
\includegraphics[width=8cm,height=5cm]{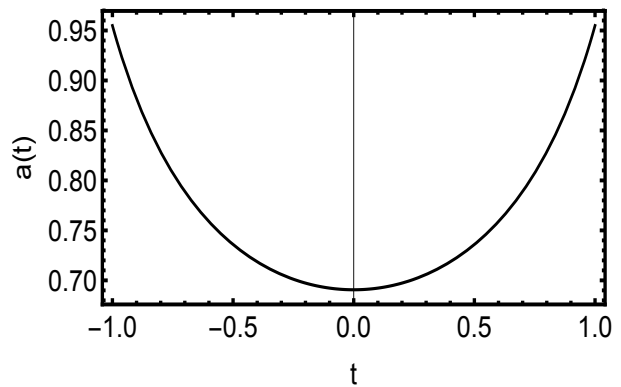}
\caption{  {\em The evolution of the scale factor $a(t)$ that is generated by 
$\mathcal{K}(\phi,B)= \phi/2+(B-B_0)^2$ and $\mathcal{G}=\xi B$, with $B_0=25.5$ and 
 $\xi=0.1$ in $M_{pl}$ 
units. 
}  }
\label{scale2}
\end{figure}

Having in mind the qualitative information for the form of $f(B)$ obtained through   
the above reconstruction procedure, we can now reverse the   procedure 
and impose the form of $f(B)$ a priori, and then extract the induced scale factor.
We choose  
\begin{equation}
\mathcal{K}(\phi,B)= \phi/2+(B-B_0)^2,
\end{equation}
where $B_0$ is a constant. Substituting it into equations (\ref{FR1}), 
(\ref{FR2}) and (\ref{chiequation}), alongside with $\mathcal{G}=\xi B$, we 
result to three second order differential equations for  
$a(t)$, $\phi(t)$ and $\chi(t)$. Elaborating the system numerically we extract the scale 
factor, which is presented in  Fig. \ref{scale2}.
  Thus, we do verify that new gravitational scalar-tensor theories with   
$\mathcal{K}(\phi,B)= \phi/2+(B-B_0)^2$ and $\mathcal{G}=\xi B$ lead to the 
realization of a cosmological bounce.

Finally,  we mention that the above procedure can be repeated in the presence of 
the matter sector, and we find that the same forms of the functions 
$\mathcal{K}$ and $\mathcal{G}$  in the presence of dust matter give rise to 
a matter bounce.

\subsubsection{General conditions for a bounce}

We close this subsection by examining analytically the conditions for the bounce 
realization in the theories and models at hand, namely $H=0$ and $\dot{H}>0$ at the 
bounce point. In particular,  the first Friedmann equation \eqref{FR1} provides 
the general equation satisfied by the Hubble function, namely 
\begin{equation}
  H^2+b H+c=0,
\end{equation}
where 
\begin{eqnarray}
&&b = 
-2\dot{\phi}^3\mathcal{G}_B\\
&&c = 
-\frac{1}{6}\dot{\chi}^2+\frac{1}{12}e^{-2\sqrt{\frac23}\chi}\mathcal{K}+\frac{2}{9}\dot{
\phi }
^2(\dot{\phi}\sqrt{6}\dot{\chi}-3\ddot{\phi})\mathcal{G}_B\nonumber\\
&&\ \ \ \ \ \,  -\frac{1}{6}
e^{-\sqrt{\frac23}\chi}\left[\dot{B}\dot{\phi}\mathcal{G}_B+\frac{\phi}{2}+\dot{\phi}
^2(\mathcal{G}_{\phi}-2\mathcal{K}_B)\right].
\label{conditions}
\end{eqnarray}
The general solution of the above quadratic equation is
\begin{eqnarray}
H 
&=&\frac{-b\pm \sqrt{b^2-12c}}{6}.
 \label{general}
\end{eqnarray}
Hence, in order for the first bounce condition, namely $H_{b}=0$ (the subscript ``b'' 
denotes the value at the bounce point), to be 
satisfied, we deduce that at the bounce point we need $c=0$.
In the case of Model I above, i.e. for  $\mathcal{K}=\phi/2$ and 
$\mathcal{G}=\mathcal{G}(B)$, this condition becomes 
\begin{eqnarray}
&&
\!\!\!\!\!\!\!\!\!\!\!\!\!\!\!
-3\dot{\chi}^2
+ \frac{3\phi}{4}e^{-2\sqrt{\frac23}\chi}  +4\dot{\phi}
^2(\dot{\phi}\sqrt{6}\dot{\chi}\nonumber\\
&&-3\ddot{\phi})\mathcal{G}_B-3
e^{-\sqrt{\frac23}\chi}\left(\dot{B}\dot{\phi}\mathcal{G}_B+\frac{\phi}{2}\right)=0
\end{eqnarray}
at the bounce point,  while for Model II above, i.e. for  $\mathcal{K}=\phi/2+f(B)$ and 
$\mathcal{G}=\xi B$, this condition becomes  
\begin{eqnarray}
&&\!\!\!\!\!\!\!\!\!\!\!\!\!
-6\dot{\chi}^2
+3e^{-2\sqrt{\frac23}\chi}
\left[\frac{\phi}{2}+f(B)\right]+8\dot{
\phi}^2(\dot{\phi}\sqrt{6}\dot{\chi}\nonumber\\
&&
\    
-3\ddot{\phi})\xi-6
e^{-\sqrt{\frac23}\chi}\left(\dot{B}\dot{\phi}\xi+\frac{\phi}{2}-2\dot{\phi}^2f_B\right)=0
\end{eqnarray}
at the bounce point.
The above conditions simplify further once we consider the forms of the functions 
$\mathcal{G}(B)
$ and $f(B)$ obtained above.
 
Concerning  the second bounce condition, namely  $\dot{H}>0$ at the bounce point, 
from Eq. \eqref{FR2} we acquire
\begin{equation}
2\dot{\chi}^2+e^{-2\sqrt{\frac23}\chi}\mathcal{K}+2
e^{-\sqrt{\frac23}\chi}\left(\dot{B}\dot{\phi}\mathcal{G}_B-\frac{\phi}{2}+\dot{\phi}
^2\mathcal
{G}_{\phi}\right)>0.
\end{equation}
Thus, in the cases of Model I and Model II respectively we obtain
\begin{eqnarray}
4\dot{\chi}^2+ e^{-2\sqrt{\frac23}\chi}\phi+4
e^{-\sqrt{\frac23}
\chi}\left(\dot{B}\dot{\phi}\mathcal{G}_B-\frac{\phi}{2}\right)>0 
\end{eqnarray}
and
\begin{equation}
2\dot{\chi}^2+ e^{-2\sqrt{\frac23}\chi}\left[\frac{\phi}{2}+f(B)\right]+2
e^{-\sqrt{\frac23}\chi}\left(\dot{B}\dot{\phi}\xi-\frac{\phi}{2}\right)>0.
\end{equation}

We mention here that in the analysis of the previous paragraphs we took the above general 
requirements into account in order to determine the initial conditions for the 
differential equations at the bounce point. Indeed, as we mentioned above, in order to be 
able to satisfy the above requirements and obtain a bounce, one must go beyond the simple 
$\mathcal{K}$ and $\mathcal{G}$ forms of \cite{Saridakis:2016ahq} that
were adequate to describe the late-time universe.

\subsection{Reconstruction of cyclic evolution}

In this subsection we extend the above analysis in order to construct a sequence of 
bounces and turnarounds, namely in order to obtain a cyclic cosmological evolution.
As a first step we will consider a specific cyclic scale factor and we will reconstruct 
the corresponding $\mathcal{K}$ and $\mathcal{G}$ forms that generate it. Then, 
having obtained information for their qualitative behavior, we will consider specific 
$\mathcal{K}$ and $\mathcal{G}$ forms and show that they lead to cyclic evolution. In the 
following paragraphs we examine two such cases separately.

\subsubsection{Model I: $\mathcal{K}=\phi/2$ and $\mathcal{G}=\mathcal{G}(B)$}

As a first model we consider $\mathcal{K}=\phi/2$ and $\mathcal{G}=\mathcal{G}(B)$. 
We start considering an oscillating scale factor of the form  
\begin{equation}
a(t)=A\sin(\omega t)+a_c,
\label{cyclicscalefactor}
\end{equation}
with $a_{c}-A>0$ the scale factor  at the bounce point, and  $A+a_c$ the scale 
factor value at the turnaround.
Inserting these    into  
(\ref{FR1}),(\ref{chiequation}) and (\ref{phiequation}), 
and replacing 
$\mathcal{G}_B(t)=\mathcal{G}'(t)/B'(t)$, and 
$\mathcal{G}_{BB}(t)=(\mathcal{G}''(t)B'(t)-B''(t)\mathcal{G}')/(B'(t))^3$,
we obtain a system of differential 
equations for $\phi(t)$, $\chi(t)$ and $G(t)$. We mention that since cyclic cosmology 
describes the whole universe evolution, and not only the phase around the bounce, we must 
necessarily include the matter sector. Without loss of generality we focus on the dust 
matter case, where $p_m=0$ and thus the continuity equation (\ref{rhoevol}) leads to 
$ \rho_m=\rho_{mb}(a_c-A)^3/a^3$.

\begin{figure}[ht]
\centering
\includegraphics[width=8cm,height=5cm]{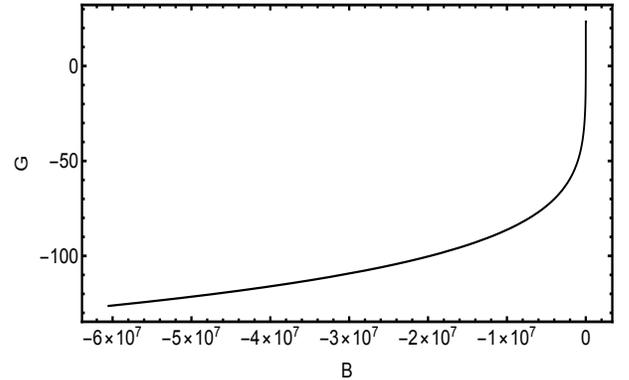}
\caption{\small  {\em The reconstructed $\mathcal{G}(B)$ that generates the cyclic 
scale 
factor (\ref{cyclicscalefactor}), in the case where 
$\mathcal{K}=\phi/2$. The parameters have been chosen as  $ A=0.05$,
$ a_c=10$, $\omega=1$. All quantities are measured in $M_{pl}$ units.
}  }
\label{GBcyclic}
\end{figure}

Since the above system of differential equations cannot be 
solved analytically, we elaborate it numerically and we obtain the solutions for 
$\phi(t)$, $\chi(t)$ and $G(t)$, and thus  for $B(t)$ too, and therefore 
 we acquire the function $\mathcal{G}(B)$ in a parametric form.
In Fig. \ref{GBcyclic} we depict this reconstructed $\mathcal{G}(B)$, which is the one 
that gives rise to the cyclic scale factor (\ref{matterbounce}).

\begin{figure}[ht]
\centering
\includegraphics[width=8cm,height=5cm]{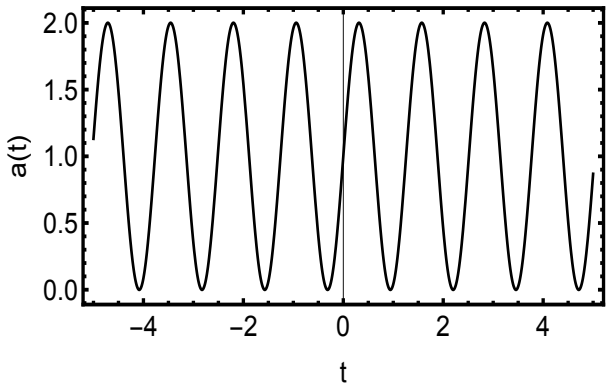}
\caption{  {\em The evolution of the scale factor $a(t)$  that is generated by   
the exponential form $\mathcal{G}(B)= G_0 e^{G_1 B}$, in the case where 
$\mathcal{K}=\phi/2$, with $G_0=2.4$ and $G_1=0.8$ in $M_{pl}$ 
units. 
}  }
\label{atcyclicI}
\end{figure}

As we observe from the above procedure, and in particular from  Fig. \ref{GBcyclic},
in order to acquire a cyclic scale factor in the 
case where  $\mathcal{K}=\phi/2$, we need a $\mathcal{G}(B)$ function that resembles 
an exponential function of $B$. Thus, we can now reverse the reconstruction procedure and 
impose an exponential form of 
$\mathcal{G}(B)$, namely
\begin{equation}
\mathcal{G}(B)= G_0 e^{G_1 B},
\label{Gform1}
\end{equation}
where $G_0$ and $G_1$ are parameters. Substituting it into (\ref{FR1}), 
(\ref{FR2}) and (\ref{chiequation}), with $\mathcal{K}=\phi/2$, we 
obtain three second-order differential equations for  
$a(t)$, $\phi(t)$ and $\chi(t)$. Elaborating the system numerically we extract the scale 
factor, and we present it in Fig. \ref{atcyclicI}.
Thus, we can see that new gravitational scalar-tensor theories with   
$\mathcal{K}=\phi/2$ and $\mathcal{G}(B)=G_0 e^{G_1 B}$ can indeed produce a cyclic 
universe. We mention here that the exponential $\mathcal{G}(B)$ may also lead to 
a single bounce realization, as we saw in the previous subsection, however what 
distinguishes the two possibilities are the parameter values.

\subsubsection{ Model II: $\mathcal{K}=\phi/2+f(B)$ and $\mathcal{G}=\xi B$}

\begin{figure}[ht]
\centering
\includegraphics[width=8cm,height=5cm]{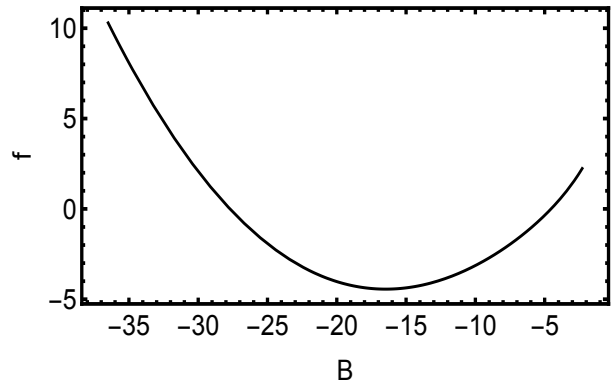}
\caption{\small  {\em The reconstructed $f(B)$  that generates the cyclic 
scale factor (\ref{cyclicscalefactor}), in the case where 
 $\mathcal{K}=\phi/2+f(B)$ and $\mathcal{G}=\xi B$. The parameters have been chosen as  $ 
A=0.05$,
$ a_c=10$, $\omega=1$ and $\xi=0.1$. All quantities are measured in $M_{pl}$ units.
}  }
\label{fBcyclic}
\end{figure}

We now study cyclicity in a different model, namely in the case where  
$\mathcal{K}=\phi/2+f(B)$ and $\mathcal{G}=\xi B$, with $\xi$ a parameter.
Similarly to the previous paragraph, we first consider the cyclic scale 
factor (\ref{cyclicscalefactor}), in order to numerically reconstruct  $f(B)$ and acquire 
a qualitative picture of its form. Substituting these  into equations
(\ref{FR1}),(\ref{chiequation}) and (\ref{phiequation}), replacing  
$\mathcal{K}_B=f'(t)/B'(t)$ and 
$\mathcal{K}_{BB}(t)=(f''(t)B'(t)-B''(t)f')/(B'(t))^3$, and considering a dust matter 
sector, we obtain a system of differential equations for $\phi(t)$, $\chi(t)$ and $f(t)$.
Solving it numerically we extract $\phi(t)$, $\chi(t)$ and $f(t)$, and therefore $B(t)$ 
too, and thus we obtain the function $\mathcal{G}(B)$ in a parametric form.
In Fig. \ref{fBcyclic} we show this reconstructed $\mathcal{G}(B)$, which is the one 
that gives rise to the cyclic scale factor (\ref{matterbounce}).

\begin{figure}[ht]
\centering
\includegraphics[width=8cm,height=5cm]{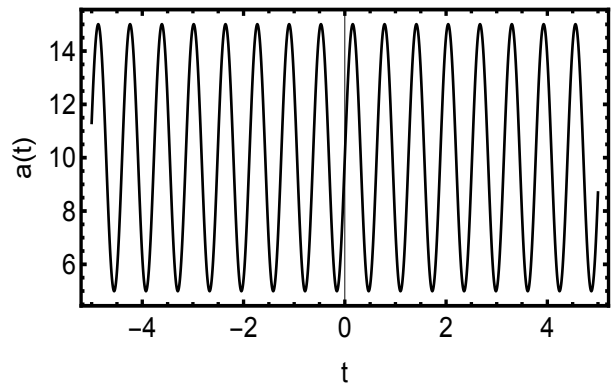}
\caption{  {\em The evolution of the scale factor $a(t)$ that is generated by 
$\mathcal{K}(\phi,B)= \phi/2+(B-B_0)^2$ and $\mathcal{G}=\xi B$, with $B_0=19.7$ and 
 $\xi=0.1$ in $M_{pl}$ 
units. 
}  }
\label{atcyclicII}
\end{figure}

As we observe from Fig. \ref{fBcyclic}, in order to acquire a cyclic scale factor in the 
case where  $\mathcal{K}=\phi/2+f(B)$ and $\mathcal{G}=\xi B$, we need a  $f(B)$
function that resembles an parabolic function. Having this in mind we can now  consider 
as an input the parabolic form
\begin{equation}
\mathcal{K}(\phi,B)= \phi/2+(B-B_0)^2,
\end{equation}
with $B_0$ a constant. Inserting into (\ref{FR1}), 
(\ref{FR2}) and (\ref{chiequation}), with $\mathcal{G}=\xi B$, we obtain
a system of differential equations for  
$a(t)$, $\phi(t)$ and $\chi(t)$. Solving the equations numerically we extract the scale 
factor, and we depict it in Fig. \ref{atcyclicII}.
Hence, we deduce that the theories at hand  
with   
$\mathcal{K}(\phi,B)= \phi/2+(B-B_0)^2$ and $\mathcal{G}=\xi B$ can indeed produce a 
cyclic universe. Note that, as we saw in the previous subsection, this parabolic form 
for $\mathcal{K}(\phi,B)$ may also lead to a single bounce solution, 
however what distinguishes the two possibilities are the parameter values.

\section{Conclusions}
\label{Conclusions}

In this work we investigated the bounce and cyclicity realization in the framework of new 
gravitational scalar-tensor theories. In particular, in these theories one considers a 
Lagrangian with 
the Ricci scalar and its first and second derivatives, however in 
a specific combination that makes the theory free of ghosts. Transforming into the 
Einstein frame, one can show that these constructions propagate  $2+2$ degrees of 
freedom, and thus they are a subclass of bi-scalar extensions of general relativity. 
Nevertheless, the fact that these theories can be  expressed in pure 
geometrical terms is a significant advantage.

We studied bouncing and cyclic solutions in various cases, reconstructing the forms of 
the functions $\mathcal{K}(\phi,B)$ and $\mathcal{G}(\phi,B)$ that can give rise to a 
given scale factor. Thus, having in mind the necessary qualitative form of these 
functions, we were able to reverse the procedure in the more physical base, namely we 
considered suitable simple functions $\mathcal{K}(\phi,B)$ and $\mathcal{G}(\phi,B)$ 
exhibiting this qualitative form, and we showed that they can give rise to a bounce or 
cyclic scale factor. 
 
We close this work by referring to the perturbations of the obtained background 
solutions. In every bouncing scenario the analysis of   perturbations is 
necessary, since they  are related to observables such as the tensor-to-scalar ratio. 
While in inflationary cosmology the generation of the primordial power spectrum requires 
that  the cosmological fluctuations emerge initially  inside the Hubble horizon, then 
they exit it, and later on they  re-enter, in a bounce scenario the situation is 
radically different. In particular, in bouncing cosmology  the quantum
fluctuations around the initial vacuum state are generated in the contraction phase 
before the bounce, they exit the Hubble radius  as contraction continues, since the 
Hubble horizon decreases faster than the wavelengths of the primordial fluctuations, 
then the bounce happens, and finally they  re-enter inside the horizon at later times in 
the expanding phase. Definitely, at the bounce point the background evolution 
could affect the perturbations scale dependence, however one expects this effect to be 
important only in the UV regime, where the gravitational modification effects play role, 
while the IR regime, which is responsible for  the   primordial perturbations related to 
the large-scale structure, remains almost unaffected 
\cite{Battarra:2014tga,Quintin:2015rta,Koehn:2015vvy}. 

Although the generation of perturbations in bouncing models with one extra scalar 
degree of freedom is well understood and studied 
\cite{Novello:2008ra,Allen:2004vz,Qiu:2011cy}, in the case where the underlying theory has 
more than one extra scalar degrees of freedom, where both of them contribute to the 
bounce, the perturbation generation has not been studied in detail. In particular, the 
examined scenarios in this subclass assume that one of the two  extra scalar 
degrees of freedom is the dominant one at some point 
\cite{DiMarco:2002eb,Bozza:2005wn,Bozza:2005xs,Kim:2006ju,Koyama:2007mg,Cai:2013kja}. 
However, this approach cannot be straightforwardly followed in scenarios where both 
fields have more or less equal contribution to the bounce realization, and one can see 
that the present scenario lies in this category. Hence, the analysis of perturbation 
generation in the bouncing scenario at hand has to be performed in a thorough and 
systematic way, through the full and detailed perturbation generation analysis of general 
two-field bounces. For this investigation one could use concepts and techniques of the 
perturbation generation in two-field inflation  
\cite{GarciaBellido:1995qq,Wands:2002bn,Vernizzi:2006ve,Lalak:2007vi,Peterson:2010mv} 
(which is different from single-field inflation with a second sub-dominant field such as 
in cases of hybrid inflation \cite{Linde:1993cn,GarciaBellido:1996qt,Lyth:2005qk}). 
Nevertheless, this detailed analysis of perturbation generation in two-field bouncing 
models is a separate work that lies beyond the scope of the present project, and it is 
left for a future investigation.

In summary, we showed that the new gravitational scalar-tensor theories, namely a subclass 
of bi-scalar extension of general relativity that can be constructed by pure geometrical 
terms,
can naturally give rise to bouncing and cyclic behavior. This capability acts as an 
additional advantage for these theories.

\section*{Acknowledgments}
  This article is based upon work from COST Action ``Cosmology and Astrophysics Network
for Theoretical Advances and Training Actions'', supported by COST (European Cooperation
in Science and Technology).

\end{document}